**Perspectives on atmospheric evolution from noble gas and nitrogen isotopes on Earth, Mars & Venus**

**Guillaume Avice** (ORCID #0000-0003-0962-0049)[1*]

**& Bernard Marty** (ORCID #0000-0001-7936-1519)[2]

[1]Université de Paris, Institut de physique du globe de Paris, CNRS, F-75005 Paris, France

[2]Centre de Recherches Pétrographiques et Géochimiques, UMR 7358 CNRS & Université de Lorraine, 15 Rue Notre Dame des Pauvres, 54501 Vandoeuvre-les-Nancy, France

*corresponding author's email: avice@ipgp.fr

**Abstract**

The composition of an atmosphere has integrated the geological history of the entire planetary body. However, the long-term evolutions of the atmospheres of the terrestrial planets are not well documented. For Earth, there were until recently only few direct records of atmosphere's composition in the distant past, and insights came mainly from geochemical or physical proxies and/or from atmospheric models pushed back in time. Here we review innovative approaches on new terrestrial samples that led to the determination of the elemental and isotopic compositions of key geochemical tracers, namely noble gases and nitrogen. Such approaches allowed one to investigate the atmosphere's evolution through geological period of time, and to set stringent constraints on the past atmospheric pressure and on the salinity of the Archean oceans. For Mars, we review the current state of knowledge obtained from analyses of Martian meteorites, and from the direct measurements of the composition of the present-day atmosphere by rovers and spacecrafts. Based on these measurements, we explore divergent models of the Martian and Terrestrial atmospheric evolutions. For Venus, only little is known, evidencing the critical need for dedicated missions.

*keywords: atmosphere, noble gases, nitrogen, geochemistry*

**Introduction**

Planetary atmospheres have accumulated some of their volatile elements either by early impact delivery or by degassing of the planetary body over long periods of time. They also have lost some of these elements by atmospheric escape or subduction into the silicate portions of the planet. For these reasons, they constitute an integrated archive of the evolution of the entire planet. Studying the relative elemental ratios and isotopic compositions of key tracers like noble gases and nitrogen provide crucial information on the origin and evolution of planetary atmospheres. Here, we review results from recent studies on the elemental and isotopic composition of volatile elements in ancient rocks and their implications for the evolution of the Earth's atmosphere (Section 1). Some of these results are then compared with the case of Mars (Section 2). Finally, the state of (lack of) knowledge for the Venus atmosphere is briefly reviewed (Section 3).

**1. Earth's atmosphere**

**1.1. Noble gases**

Noble gases are excellent tracers of the geodynamics of a planetary system (see reviews by Moreira 2013; Ozima and Podosek 2002; Porcelli et al. 2002)). Due to their inertness, their low abundances and the production of some of their isotopes by radioactive decay of extinct (*e.g.* $^{129}$I, $t_{1/2}$=15.7 Myr) and extant (*e.g.* $^{40}$K, $t_{1/2}$ = 1.25 Gyr) radionuclides, their measurement can track physical processes such as volatile delivery during Earth's accretion, magmatic degassing, atmospheric escape etc.

**1.1.1. The long-standing xenon paradox**

Xenon (Xe), the least abundant stable noble gas, has nine stable isotopes which relative proportions reflect the origin of this element as well as processes that changed its abundance through time. Atmospheric xenon on Earth and Mars presents two intriguing features. Firstly, its abundance is lower than what would be predicted if noble gases in planetary atmospheres were following the same abundance pattern as in chondritic meteorites (Fig. 1a). In other words, the $^{132}Xe/^{84}Kr$ ratio is one order of magnitude lower than that of chondritic gases (Pepin, 1991). This depletion suggests that xenon is "missing" in the atmospheres of Earth and Mars. The only data for Xe in the Venus atmosphere gives an upper limit of 10 ppb (Wieler (2002) and refs. therein) and does not allow one to evaluate if xenon is also missing in the atmosphere of this planet. Secondly, the isotopic spectra of Earth and Mars atmospheric xenon show enrichments in heavy isotopes relative to the light ones compared to other cosmochemical Xe components (Fig. 1b) such as Xe in the Q phase, the major carrier of noble gases in chondrites (Busemann et al. 2000), Xe in the solar gas (SW-Xe) from which the solar wind (SW) originates (Meshik et al. 2014) or U-Xe the progenitor of atmospheric xenon (Pepin 1991). The strong mass-dependent fractionation of Xe isotopes (3-4% per u, i.e., atomic mass unit) is the second feature of the so-called "xenon paradox": atmospheric xenon, the heaviest noble gas, and thus the least prone among noble gases to escape by mass-dependent thermal processes, is depleted in abundance and isotopically fractionated in favor of heavy isotopes in both Earth and Mars atmospheres.

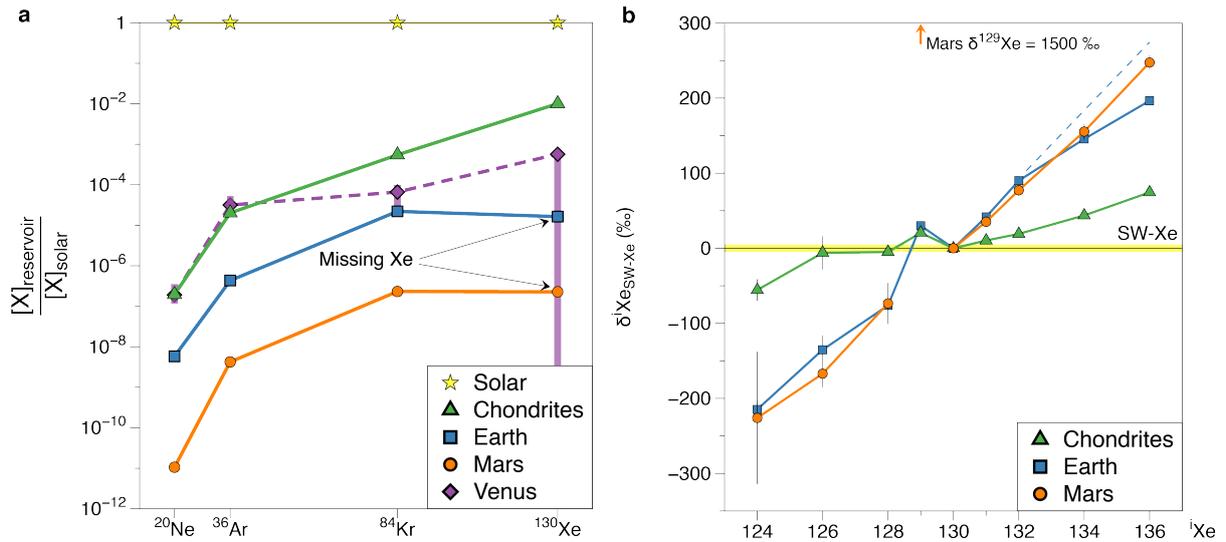

*Figure 1: Illustration of the xenon paradox.* (a) *Abundances of noble gases in planetary atmospheres and in chondritic meteorites normalized to the solar composition. Modified from* Dauphas and Morbidelli (2014). *Data for Venus are from Wieler (2002). (b) Isotopic composition of Xe in chondrites* (Busemann et al. 2000) *and Earth* (Ozima and Podosek 2002) *and Mars* (Mathew et al. 1998) *atmospheres. Isotopic compositions are given using the delta notation normalized to $^{130}Xe$ and to the isotopic composition of Solar Wind Xe (SW-Xe, $\delta^i Xe_{SW-Xe} = 0$ ‰): $\delta^i Xe_{SW-Xe} = 1000 \times \left( \frac{(^iXe/^{130}Xe)_{component}}{(^iXe/^{130}Xe)_{SW-Xe}} - 1 \right)$. The dashed line illustrates the underabundance of $^{134}Xe$ and $^{136}Xe$ relative to mass-dependent fractionation of SW-Xe.*

Previous theoretical studies (Pepin 1991; Shcheka and Keppler 2012; Tolstikhin and O'Nions 1996) explained this paradox by early processing of noble gases during Earth's forming events (see also the review by Dauphas and Morbidelli (2014)). Overall, these models assume separation of noble gases by magmatic/petrological processes, storage of these but xenon in solid Earth reservoirs (*e.g.*, core, mantle or magma ocean), escape of Xe from the early atmosphere and fractionation of its isotopes, and subsequent degassing of the other noble gases. These models were somehow *ad hoc* and require a suite of complex, fine-tuned events. Recent experimental studies propose that missing Xe might be stored inside silicate structures in the deep Earth (Crépisson et al. 2018; Crépisson et al. 2019; Sanloup 2005). However, a selective retention of Xe inside the Earth could only explain the elemental depletion of atmospheric Xe, but not the large isotopic fractionation relative to cosmochemical end-members. Additionally, storage of atmospheric Xe, which is ultimately derived from U-Xe (see next paragraph), in the deep Earth is not consistent with a chondritic origin of mantle Xe (Caracausi et al. 2016; Péron and Moreira 2018).

There is an additional remarkable feature of atmospheric Xe aside of its elemental depletion and isotopic fractionation. Isotopic fractionation of atmospheric Xe follows a mass-dependent law in which the extent of fractionation between two isotopes relates to the mass difference between these two isotopes. This law applies to almost all Xe isotopes (except for a mono-isotopic excess of $^{129}$Xe due to a nuclear effect, the decay of extinct $^{129}$I). However, the two heaviest Xe isotopes, $^{134}$Xe and $^{136}$Xe, depart from this law and are under-abundant with respect to other Xe isotopes (Pepin 1991). This Xe component has been nicknamed U-Xe and has resisted decades of investigation (Pepin 1994). Recently, the analysis of noble gases released by Comet 67P/Churyumov-Gerasimenko by the ROSINA instrument onboard of the Rosetta spacecraft (Balsiger et al. 2015; Marty et al. 2017; Rubin et al. 2018) revealed that cometary Xe is depleted in these two $^{134}$Xe and $^{136}$Xe isotopes, corresponding to a depletion in xenon nuclides created by the r-process of stellar nucleosynthesis. This feature, shared with U-Xe, suggests the persistence of large-scale nucleosynthetic heterogeneities across the Solar System (Avice et al. 2020). Mass balance calculations show that the addition of ~20 % cometary Xe to meteoritic Xe accounts for the peculiar composition of U-Xe for $^{128,130-136}$Xe isotopes (Marty et al. 2017). The interior of the Earth contains meteoritic Kr and Xe (Holland et al. 2009; Péron and Moreira 2018), implying that the addition of cometary xenon to the Earth's atmosphere took place when the terrestrial mantle, or at least some of its regions, were already isolated from contribution of atmospheric volatiles. Thus, cometary contributions appear to have been late with respect to the main phases of terrestrial accretion. In summary, xenon in the terrestrial atmosphere originates from a peculiar component, U-Xe, that may consist in a mixture of cometary and meteoritic gases. This mixture is likely to have occurred during "late" contribution of material from the outer solar system, by the end of the Earth's building stage (O'Brien et al. 2018). Once formed, this component has been isotopically fractionated by losing light Xe isotopes with respect to heavy ones. The analysis of ancient atmospheric xenon, reported in the next section give clues about this process.

**1.1.2. Evolution of the isotopic composition of Earth's atmospheric Xe: data and models**

Recent measurements of paleo-atmospheric gases trapped in fluid inclusions, in lattices of ancient minerals, in carbon-rich lithologies or in free fluids collected underground reveal that the isotopic composition of atmospheric xenon evolved through time (Fig. 2a). The analysis of Archean samples show isotopic compositions of past-atmospheric Xe intermediate between U-Xe and modern Xe (Fig. 2b), demonstrating that Xe isotopic fractionation did not take place during the Earth's building stages but was progressive through time (Avice et al. 2017; Avice et al. 2018b; Bekaert et al. 2018; Bekaert et al. 2019; Pujol et al. 2011). This evolution corresponds to a progressive enrichment of atmospheric Xe in heavy isotopes relative to the light ones (Fig. 2b), presumably initiating from U-Xe at 4.5 Ga and reaching the modern value around 2 Ga ago. In contrast, there is no evidence for a change during the last 3.5 Gyr of the isotopic compositions of neon, argon (for its non-radiogenic isotopes) and krypton (Avice et al., 2018 and refs. therein).

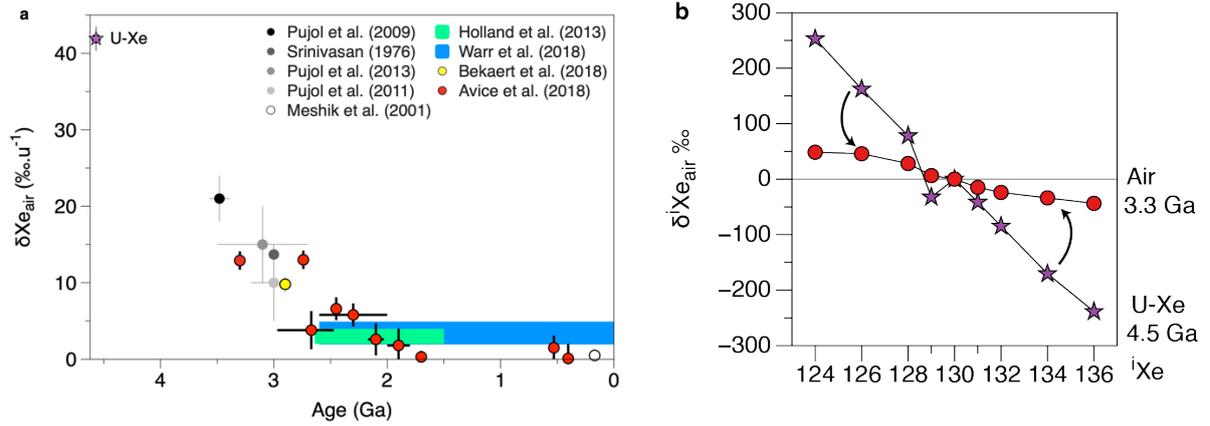

*Figure 2: Evolution of atmospheric Xe.* (a) *Evolution of the isotopic composition of atmospheric Xe with time. Most data are from analyses of gases trapped in mineral samples* (Avice et al. 2018b; Bekaert et al. 2018; Holland et al. 2013; Meshik et al. 2001; Pujol et al. 2013; Pujol et al. 2009; Srinivasan 1976). *References to* Holland et al. (2013) *and* Warr et al. 2018) *correspond to measurements of gases contained in ancient fluids collected underground. Error bars correspond to 1σ.* (b) *Isotopic composition of atmospheric Xe at 4.5 Ga (U-Xe) and 3.3 Ga (Barberton,* Avice et al. 2017)). *Isotopic compositions are given using the delta notation normalized to $^{130}$Xe and to the isotopic composition of Earth atmospheric Xe ($\delta^i Xe_{air}$= 0 ‰): $\delta^i Xe_{air} = 1000 \times (\frac{(^i Xe/^{130}Xe)_{sample}}{(^i Xe/^{130}Xe)_{air}} - 1)$. Curved arrows depict the progressive evolution of the isotopic composition of atmospheric Xe.*

Even if the temporal resolution of the dataset remains low due to the difficulties to date paleo-atmospheric signals (Avice et al. 2017; Pujol et al. 2013; Warr et al. 2018; Bekaert et al. 2019), some general features of the evolution of Xe are observed. Firstly, the evolution of the isotopic composition of atmospheric Xe is a long-term process covering several Gyr. At 3.5 Ga, only half of the isotopic fractionation was achieved. This suggests that the isotopic composition of atmospheric Xe cannot be solely explained by early episodes of vigorous hydrodynamic escape (Pepin 1991; Tolstikhin and O'Nions 1996). Secondly, the evolution was progressive without retrograde evolution. These observations are consistent with a transfer of atmospheric xenon, concomitant with its isotopic fractionation, from the atmosphere to another reservoir. Because mantle Xe appears to be a mixture of modern Xe and meteoritic Xe, it cannot constitute the complementary reservoir, and the most likely solution is a time-dependent escape of atmospheric xenon from the atmosphere to space. Such escape had to be specific to Xe. Since it did not affect the other noble gases, it cannot have resulted from standard thermal (mass-dependent) escape processes. This supports a process where part of Xe escaped from the atmosphere and left the remaining fraction isotopically fractionated. Some pause in the isotopic fractionation of atmospheric Xe might have existed (see for example data of Avice et al. (2018b) between 3.3 and 2.7 Ga) but remain to be confirmed (Bekaert et al. 2018). Finally, the evolution appears to have ended between 2.5 and 2.0 Ga. Even if the exact timing could be determined by future studies with higher temporal and isotopic resolutions, this temporal sequence overlaps with a major change in the chemistry of the Earth's atmosphere, the Great Oxidation Event. The facts that (i) the modern isotopic composition was attained only ~2 Ga ago and (ii) mantle-derived Xe does not show evidence of an Archean fractionated Xe component in the mantle demonstrate that Xe was not efficiently subducted and retained in the Earth's mantle before 2.5 Ga (Avice et al. 2017; Parai and Mukhopadhyay 2018; Péron and Moreira 2018).

**1.1.3. Escape models and research avenues**

Geochemical data obtained on paleo-atmospheric samples show a progressive evolution of the isotopic composition of atmospheric Xe. However, until recently, the physical mechanism behind this trend remained elusive. Some studies suggested a selective long-term escape and isotopic fractionation of atmospheric Xe. A model was first proposed by Hébrard and Marty (2014) for which, during the Archean, a fraction of atmospheric Xe was ionized and trapped in organic haze in the Archean atmosphere. Xenon was preferentially ionized and trapped in organic haze relative to other noble gases (e.g., Kr), for two cumulative reasons. First, the ionization potential of xenon (12.13 eV) is the lowest among noble gases (> 14 eV) and other atmospheric species, e.g., $CO_2$ (13.8 eV), CO (14 eV), and $N_2$ (15.6 eV). Second, the 1D model for the Archean atmosphere developed by these authors showed that the atmospheric height at which haze production from $CH_4$ took place coincided with that of maximum Xe ionization. Trapping of xenon ions in organics results in isotopic fractionation of Xe in favor of the heavy isotopes in the trapped phase relative to the starting gas (Kuga et al. 2017; Marrocchi et al. 2011), with a fractionation factor around 1 ‰/u. The escape of the non-trapped fraction and the subsequent release of the trapped and isotopically fractionated Xe would lead to progressive depletion and isotopic fractionation of atmospheric Xe. One caveat of this model was the absence of a physical explanation for Xe escape. More recently Zahnle et al. (2019) proposed a physical framework for selective Xe escape and isotopic fractionation. The basis of this model is that Xe ions are dragged by escaping hydrogen (H) ions from the Archean atmosphere. Because of the low ionization potential of this element, Xe ions are thus likely to exist in the presence of hydrogen ions and can persist in the atmosphere due to their low recombination coefficient with oxygen, especially during the absence of oxygen in the atmosphere before the G.O.E. Coulomb interactions between Xe and H ions permitted the escaping $H^+$ to lift $Xe^+$. Even if a magnetic field was probably present and protected the atmosphere (Tarduno et al. 2014), Xe ions could have still escaped along open magnetic field lines at the magnetic poles. Hydrogen escape could have being operant on Earth until 2.0 Ga as a driving mechanism for the progressive oxidation of the Earth's surface reservoirs (Zahnle et al. 2013). If the model of Zahnle et al. (2019) is correct, the evolution of the isotopic composition of atmospheric Xe would thus be an indirect but powerful tracer of the evolution of the composition of the Earth's atmosphere. See also Catling & Zahnle (2020) for a review on the Archean atmosphere and on the potential meanings of the isotopic evolution of atmospheric xenon. The model of Zahnle et al. (2019), through escape of Xe ions together with hydrogen ions, requires a source of hydrogen. For Xe to escape, the total $H_2$ mixing ratio had to be at least 1%. If hydrogen was only in $H_2$ molecules, this value seems to be an upper limit for the Archean atmosphere (Kadoya and Catling 2019). If hydrogen originated from the photo-dissociation of H₂O vapor, its escape could have resulted in change on the volume and composition of the hydrosphere. Zahnle et al. (2019) suggested that a volume of water corresponding to about 30 % of the present oceans could have escaped although this amount would be overestimated if Xe escape happened in short and intense episodes. Alternatively, hydrogen could have originated from CH$_4$ dissociation (Catling et al. 2001). Hydrogen escape has the potential to increase the D/H ratio of hydrogen remaining in the atmosphere. This enrichment would be of the order of 25% (Zahnle et al. 1990; Zahnle et al. 2019), which is of the same order of magnitude as the difference between bulk Earth's D/H and some D/H values for chondritic meteorites (Marty et al., 2016).

**1.2. Other noble gases**

There is no other evidence for a secular evolution of other noble gases except for a gradual increase of the abundance of $^{40}$Ar in the Earth's atmosphere (Cadogan 1977; Pujol et al. 2013; Stuart et al. 2016). $^{40}$Ar is continuously produced in Earth's silicate reservoirs by radioactive decay of $^{40}$K ($T_{1/2}$ = 1.25 Gyr) and, as an atmophile element, subsequently degassed into the atmosphere. Potassium is an incompatible element and concentrates in the continental crust produced by differentiation of mafic lithologies yielding felsic compositions. This behavior implies a link between the abundance of $^{40}$Ar in the atmosphere and the rate of growth of continental crust (Allègre et al. 1987). Paleo-atmospheric data on the $^{40}$Ar/$^{36}$Ar ratios coupled with box models have been used to put constraints on the volume of continental crust with time (Cadogan 1977; Pujol et al. 2013; Stuart et al. 2016). Overall, around 75% of the volume of the continental crust was already formed at 2.5 Ga. The rate of crustal growth was lower for the last 2.5 Gyr. These results are in broad agreement with other recent crustal growth models based on hafnium isotopes in zircons (*e.g.* Dhuime et al. 2012).

## 1.3. Paleo-atmospheric dinitrogen

The $N_2$ molecule, because of its triple covalent bond, is particularly stable and $N_2$ is sometimes called the "6th noble gas" due to its relative inertness. Under oxidizing conditions, with an oxygen fugacity higher than the iron-wüstite buffer, $N_2$ is incompatible during partial melting (Mikhail and Sverjensky 2014). It readily degasses from basaltic magmas due to its low solubility, comparable to that of argon (Libourel et al. 2003). Today, $N_2$ is the most abundant molecule in the Earth's atmosphere, and was certainly an important player in the distant past (*e.g.* Catling & Zahnle 2020). Nitrogen is also a key chemical element in biomolecules (DNA, RNA, proteins) and its bio-mediated speciation allows exchanges between the biosphere and the geosphere. Atmospheric dinitrogen is reduced to $NH_3$ by bacteria carrying the nitrogenase enzyme and nitrogen is further processed to become incorporated into organic compounds. Organic N is converted to ammonium by bacteria or fungi. $NH_4^+$ ions can substitute to $K^+$ in soils and sediments. This is one major pathway to incorporate nitrogen into the rock cycle. Sedimentary N can go back in the atmosphere as $N_2$ during metamorphism induced by continental collision and in subduction zones. Oxidation of reduced N can also take place when specific bacteria convert ammonia to nitrites or nitrates. Nitrates are dissolved in groundwater or oceanic water, where denitrification can take place converting nitrates back to $N_2$. Several other reactions such as anamox, direct conversion of nitrite and ammonia to $N_2$ permits the liberation of reduced and oxidized N back to the atmosphere. Further details about the global bio-geochemical cycle of nitrogen are given by Stüeken et al. (Chapter 9).

Overall, nitrogen is exchanged between the interior of the Earth and its surface as the result of mantle degassing at ridges and plumes, the development of the continental crust storing ammonium, and subduction permitting the return of nitrogen trapped in oceanic crust back to the mantle. The cycle of nitrogen probably varied a lot throughout the course of Earth's history due to major changes such as the colonization of the oceans by bacteria, the onset and production of continental crust, the onset of subduction, the Great Oxidation Event etc. Current views on the evolution of the nitrogen cycle are debated (*e.g.* Johnson and Goldblatt 2018; Zerkle 2018)) but a record of these variations might have been kept in ancient sedimentary rocks.

Atmospheric $N_2$ has also the potential to have recorded past atmospheric escape processes (Lichtenegger et al. 2010). Contrary to the case of Mars (McElroy et al. 1976) dinitrogen is presently not escaping from the terrestrial atmosphere. The terrestrial exobase (the height at which molecules do not interact each other and can escape to space if they have sufficient velocity) extends at altitudes a few hundreds of km, whereas the magnetosphere expands up to several tens of thousands km. Hence, atmospheric species are shielded from interaction with extraterrestrial charged particles like those of the solar wind, which is a major process to escape nitrogen in the case of Mars. In the distant past, the Earth's atmosphere might not have been well shielded when the terrestrial magnetic field was weaker. Note that the shielding efficiency of planetary magnetic fields against atmospheric escape remains debated (*e.g.* Gunell et al. 2018). The altitude of the exobase also strongly depends on the composition of the atmosphere (Tian et al. 2008; Johnstone et al. 2018), which is likely to have varied. The flux of charged particles from space might also have been much higher, than today, like during exceptional solar flare events which are observed on stars younger than the Sun. Without a magnetic field, the ancient atmosphere might have lost $N_2$ in a few million years, unless some $CO_2$ in excess of the present concentration by at least two orders of magnitude would have been present to allow cooling and shrinking of the thermosphere (Lichtenegger et al. 2010).

On Mars, atmospheric nitrogen is enriched in its rare isotope $^{15}N$ relative to abundant $^{14}N$ by 60 % compared to the Martian mantle, meteoritic, and terrestrial (atmosphere, mantle) nitrogen (McElroy et al. 1976). Although $^{15}$N-rich cometary contributions cannot be dismissed (Marty et al. 2016) such enrichment is generally attributed to isotope fractionation during non-thermal (charge dependent) escape to space (McElroy et al. 1976). On Earth, atmospheric nitrogen is enriched in the heavy isotope $^{15}$N by ~ ≥ 0.5 ‰, possibly up to 3 ‰, compared to mantle N, an enrichment that could have resulted from past nitrogen escape processes (Tolstikhin & Marty 1998). From a modeling point of view, it is difficult to decipher if atmospheric escape could have affected nitrogen in the distant past, because the interplay between potentially responsible processes such as a reduced magnetosphere, and/or an expanded exobase, are model-dependent (Lichtenegger et al. 2010) and not constrained by external observations. For these reasons, it is utterly important to determine the partial pressure and the isotope composition of atmospheric nitrogen in the distant past. Several approaches have been attempted, which converge towards coherent results. They are described in the following paragraphs.

Constraints on the total paleo-atmospheric pressure have been obtained from the morphological analysis of fossil raindrops imprinted in the 2.7 Ga volcanic tuffs of the Ventersdorp Supergroup, South Africa (Som et al. 2012). The rationale is that the maximum size a falling raindrop can reach before fragmentation is a direct function of the ambient atmospheric pressure. Assuming that the studied rocks were emplaced at the sea level, the distribution of fossil raindrops, together with analogic experiments aimed at reproducing this distribution in the laboratory allowed Som et al. (2012) to propose that the total atmospheric pressure could not have been larger than two times the modern atmospheric pressure, and more probably comparable to, or lower than, the latter. However, the approach by Som et al. (2012) was subsequently criticized by Kavanagh and Goldblatt (2015) who claimed that the pressure determined by the "fossil raindrop method" is likely to be uncertain within one order of magnitude. This criticism was later refuted by Goosmann et al. (2018). More recently, Som et al. (2016) attempted to estimate the absolute Archean barometric pressure using the size distribution of gas bubbles in lava flows that solidified at sea level 2.7 Ga ago in the Pilbara craton, Australia. They proposed that the atmospheric pressure was low, about 0.23 bar.

A different approach was proposed by Marty et al. (2013) and Avice et al. (2018b) (see also Nishizawa et al. (2007)). They analyzed the elemental and isotopic compositions of noble gases and nitrogen trapped in fluid inclusions in hydrothermal quartz by vacuum crushing and static mass spectrometry. The samples were from 3.5 Ga-old Apex and Dresser formations, Pilbara Craton, Australia, and from the meso-Archean Barberton belt, South Africa. Ar-Ar dating constrained ages of the trapped fluids to be $\geq 3.0$ Ga (Marty et al. 2013) and 3.3 Ga (Avice et al. 2018), respectively. By doing sequential crushing, these authors found that data define correlations in a $N_2/^{36}Ar$ vs. $^{40}Ar/^{36}Ar$ space interpreted as a mixing between a hydrothermal end-member rich in crustal N and radiogenic $^{40}Ar$, and a low $^{40}Ar/^{36}Ar$, low $N_2/^{36}Ar$ end-member. The latter had $^{40}Ar/^{36}Ar$ ratios comparable to that of the Archean atmosphere, and was considered to represent atmospheric noble gases dissolved in Archean seawater (Fig. 3). The $N_2/^{36}Ar$ ratio of this end-member corresponds to dissolved atmospheric $N_2$ and $^{36}Ar$. Measurements of different samples led to distinct correlations with variable slopes reflecting different compositions for the hydrothermal end member. However, the correlations converged toward a common low $^{40}Ar/^{36}Ar$, $N_2/^{36}Ar$ composition. Atmospheric $N_2$ and Ar dissolve in water with known solubilities. The dissolved $N_2/^{36}Ar$ ratio is thus directly representative of the atmospheric ratio. One assumption is that the abundance of atmospheric $^{36}Ar$ did not vary with time, which is comforted by the fact that the atmospheric $^{36}Ar/^{38}Ar$ ratio is similar to that of primitive meteorites (Ozima and Podosek 2002); and by the fact that there is no known process allowing the escape of neutral Ar atoms. The partial pressure of $N_2$ can thus be deduced from the $N_2/^{36}Ar$ ratio of the Archean water end-member. Marty et al. (2013) proposed an Archean $P_{N2}$ in the range of 0.5-1 bar at $\geq$ 3 Ga, and Avice et al. (2018) confirmed that at 3.3 Ga, the Archean $P_{N2}$ was lower than the modern one.

Altogether, these studies suggest that the partial pressure of paleo-atmospheric $N_2$ was comparable to, or lower than, the modern one (see also a compilation provided by Silverman et al. (2018)). These results do not support the suggestion by Goldblatt et al. (2009) that the faint young Sun paradox (see Chapters 1 & 9) could be resolved if the paleo-atmospheric $P_{N2}$ was 2-3 times more than the modern one leading to an enhancement of the greenhouse effect by pressure broadening of $N_2$. Similarly, warming induced by $H_2$-$N_2$ collision-induced absorption (Wordsworth and Pierrehumbert 2013) seems excluded since it requires a paleo-atmospheric $P_{N2}$ three times higher than the modern one. Finally, a relatively low $P_{N2}$ in the past supports the view that, in the deep past, a nitrogen-dominated atmosphere could not have been stable against thermal escape fueled by high EUV fluxes (Johnstone et al. 2019).

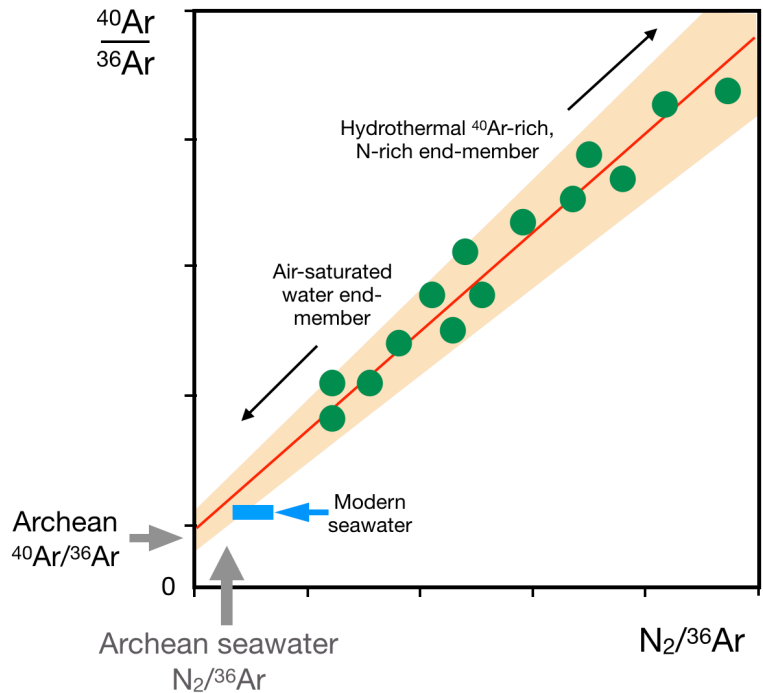

*Figure 3: Schematic illustration of the principle of the geochemical determination of the partial pressure of N₂ of the Archean atmosphere. The samples are Archean hydrothermal quartz hosting fluid inclusions. Ages of the fluids are constrained with the Ar-Ar method. Sequential heating and crushing under static high vacuum, followed by mass spectrometry analysis of Ar and N isotopes and abundances permit to obtain a series of data points for a given samples (green dots). Data define a linear relationship in a $^{40}Ar/^{36}Ar$ vs. $N_2/^{36}Ar$ diagram regarded as representing fluid mixing between a crustal component rich in radiogenic $^{40}Ar$ and in nitrogen, and a low $^{40}Ar/^{36}Ar$, $N_2/^{36}Ar$ component representing Archean sea- (or fresh-)water component, which has the Ar isotopic composition of dissolved Archean air, and a $N_2/^{36}Ar$ ratio representative of Archean air, after correction for Bunsen (solution) coefficients for both gases. As the abundance of $^{36}Ar$ is likely to have been conservative since at least the meso-Archean, the $N_2/^{36}Ar$ ratio of the Archean water can be compared directly to the modern seawater ratio. Values lower than the latter indicate $P_{N2}$ lower than today. Correlations (Marty et al. 2013; Avice et al. 2018) indicate that the Archean $P_{N2}$ was equal to, and probably lower than, the modern $P_{N2}$, thus dismissing a $P_{N2}$ three times higher than the modern one proposed to counterbalance a fainter Sun than today (Goldblatt et al. 2009).*

The nitrogen isotope composition was also investigated in the studies described above (Marty et al. 2013; Avice et al. 2018) and also in the study by Pinti et al. (2001). The $\delta^{15}N$ variations (where $\delta^{15}N = [(^{15}N/^{14}N_{sample}/^{15}N/^{14}N_{atm}) - 1] \times 1000$, in permil; ‰) can also be accounted for by mixing between a paleoseawater component having a N isotope composition similar to that of modern seawater and a crustal component having values in the range of modern crust and sediments (Fig. 4). Thus, the N isotope composition of Archean atmosphere seems to have been similar within a few permil to that of the modern atmosphere.

Taken together, these different studies converge with an atmospheric composition still dominated by $N_2$. The Archean $P_{N2}$ was at best equal to, and probably lower than, the modern total and N partial pressures. A low Archean atmospheric $N_2$ is unlikely to be the result of atmospheric N escape, since the N isotope composition did not vary. These results are somewhat unexpected for a period of time when subduction and the possibility of N recycling to the mantle might not have been operative. The Archean atmospheric N "depletion" might have been the result of nitrogen sequestration in organic matter and in the crust at a time when environment reducing conditions did not permit fixed nitrogen to be returned to the atmosphere by denitrification and oxidative weathering (Stuëken et al. 2016; Chapter 9). Combining estimates of the total atmospheric pressure with those of the $P_{N2}$ suggests that there might have been some significant amount of other atmospheric species up to 0.7 bar when uncertainties are taken into account, which leaves space to have enough greenhouse gas like $CO_2$ to counterbalance a low energy delivery by a fainter Sun at that time (Catling & Zahnle (2020) and refs. therein). The most extreme terrestrial N isotope values are about $\delta^{15}N$ = - 40 ‰ as found in some diamonds, and the convective mantle is characterized by $\delta^{15}N$ values around - 5 ‰ (Cartigny and Marty 2013). The maximum isotope fractionation of nitrogen in the terrestrial system was thus at most 40 ‰, and since 3.5 Ga it may have not varied by more than a few permil. Contrary to the case of xenon, these results suggest low nitrogen atmospheric escape since three billion years, attesting for efficient atmospheric shielding against interactions with charged particles in the upper atmosphere.

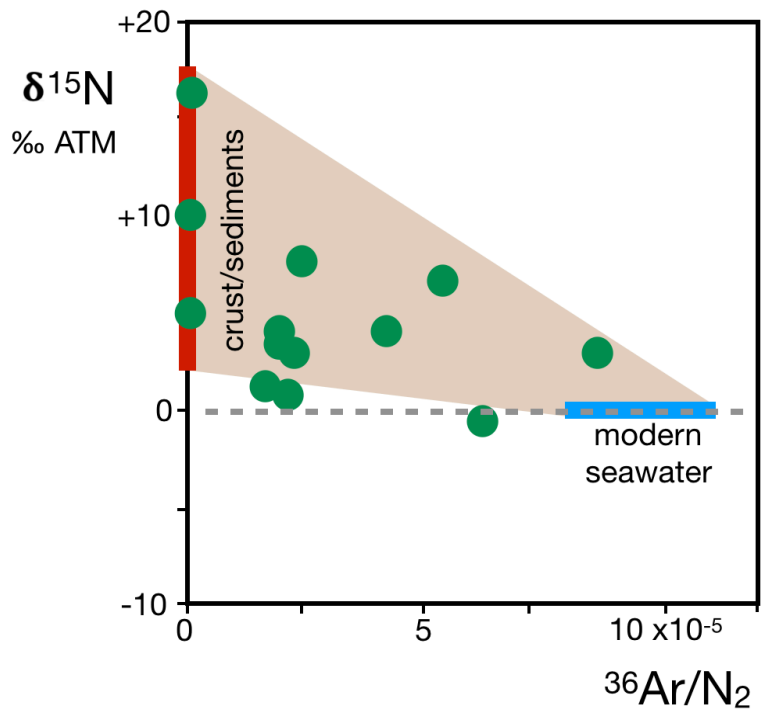

*Figure 4*: Principle of the determination of the nitrogen isotope composition of Archean atmospheric N$_2$ ($\delta^{15}$N is the $^{15}$N/$^{14}$N ratio of the sample normalized to the isotopic ratio of nitrogen in the modern atmosphere, in parts per mil - ‰). In this space, mixing between two component results in a linear relationship. Here the data (green dots) obtained from the sequential analysis of fluid inclusions trapped in Archean hydrothermal quartz rather define an area corresponding to mixing between modern seawater and the range of values typical of oceanic and continental crust as well as sediments. The Archean seawater end-member appears to have had a N isotope composition not drastically different from the modern one, within a few permil (modified from Marty et al. 2013, Avice et al. 2018 and Pinti, 2001).

Interestingly, the argon correlation approach used to determine the Archean P$_{N2}$ permitted also to constrain the salinity of the Archean oceans (Marty et al. 2018). Some of the above samples were neutron-irradiated, which synthetized Ar isotopes from Cl and K isotopes dissolved in fluid inclusions. Two end-member Cl-K-Ar isotope correlations were obtained, regarded as representing mixing between an Archean seawater component and a hydrothermal fluid end-member. Different hydrothermal end-members were identified yielding correlations with different slopes which converged towards a unique seawater end-member (Marty et al. 2018). An example of such correlation is given in (Fig. 5). The seawater end-member appears to have a Cl/$^{36}$Ar - K/$^{36}$Ar composition comparable to that of the modern oceans. Assuming again that $^{36}$Ar has been conservative in the atmosphere since 3.5 Ga, it follows that the salinity (Cl) of the oceans did not vary significantly through time. In detail, the abundance of potassium in the oceans might have been ~30% lower than today, in possible relation with a lower volume of continental crust 3.5 Ga ago. Interestingly, the equilibrium temperature for atmospheric gas dissolution appears to have been in the range 25-50°C, suggesting that the temperature of the oceans might have been higher than today, as argued (and disputed) by studies reporting oxygen/silicon isotope measurements on cherts and carbonaceous matter (*e.g.* Knauth 2005; Robert and Chaussidon 2006; Tartèse et al. 2016)). Alternatively, such temperature range could reflect local surface conditions where atmospheric gases were in equilibrium with surface water, and not global oceans' temperature. The corollary is that the halogen/seawater ratio remained constant within ~20-30% from 3.5 Ga to present, implying that the potential escape of hydrogen from photodissociation of atmospheric $H_2O$ might have been limited. A ~30% water loss is consistent with modeling of the escape of atmospheric xenon which requires approximately this amount of oceanic water to be dissociated and resulting hydrogen to be lost into space in order to lift up Xe ions (Zahnle et al. 2019).

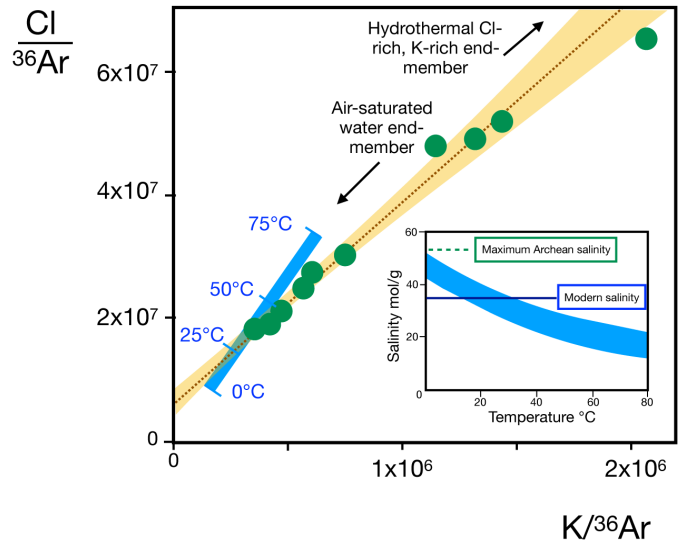

*Figure 5: Cl/$^{36}$Ar vs. K/$^{36}$Ar data (green dots) for stepwise crushing of 3.4 Ga-old Barberton sample PI02-39-2 (adapted from Marty et al. (2018)).* Data define a correlation representing mixing between an old seawater component and a hydrothermal end-member rich in crustal chlorine and potassium. Error bars are not shown here but are given in the original work. The yellow area represents the 95% confidence interval of the error-weighted correlation. The blue bar represents the composition of modern seawater for different temperatures between 0°C and 75°C. The correlation intercepts the modern composition for temperatures between 25°C and 50°C, suggesting an oceanic Cl/$^{36}$Ar ratio similar to the modern value and, for constant atmospheric $^{36}$Ar, a salinity (Cl) of the Archean oceans similar to the modern one. The insert shows the link between salinity and temperature (Marty et al. 2018). The constraint for the maximum Archean salinity is a temperature of water above the freezing point.

## 2. Martian atmosphere

The Martian atmosphere is much less dense ($P_{atm}$ = 600 Pa) than that of the Earth ($P_{atm}$ = 101,325 Pa). The major atmospheric species ($CO_2$, $N_2$) do not escape by thermal (mass-dependent) processes, at least at present. However, the isotopic compositions of hydrogen (D/H) and nitrogen are largely fractionated by several hundreds to thousands per mil, respectively (see Bogard et al. (2001) and refs therein). The two stable isotopes of argon are also isotopically fractionated, but krypton is not, and xenon is as fractionated as terrestrial xenon. Martian volatiles present an intriguing puzzle which is not resolved yet. The diversity and extent of isotope fractionation call for several types of escape processes having acted on the atmosphere of the red planet during extended periods of time. In this part we focus on among the most important isotopic effects which affected xenon and nitrogen.

**2.1. The strange case of Martian atmospheric xenon**

If the abundance and the isotopic composition of Earth atmospheric Xe result from a specific escape mechanism, it is interesting to turn ourselves toward Mars and evaluate if such a mechanism was operating on the red planet. The isotopic fractionation of terrestrial Xe is 3.6 ‰/u (computed with the atmospheric $^{128}Xe/^{130}Xe$ ratio) relative to its precursor, U-Xe (Pepin and Porcelli 2002). For Mars the precursor seems to be SW-Xe and not U-Xe, as observed in the Chassigny meteorite thought to have sampled noble gases from the Martian mantle (Ott 1988) on one hand, and from the isotopic composition of Martian atmospheric Xe after correction for mass-dependent fractionation, on another hand. The isotope fractionation of Xe in the Martian atmosphere is 3.3 ‰/u when the Xe isotope composition of the Martian atmospheric gases trapped in Martian meteorites (Swindle 2002) is considered (relative to SW-Xe (Meshik et al. 2014)). It is only 2.4 ‰/u when the direct measurement of the atmosphere by the Mars Sample Laboratory's mass spectrometer (Conrad et al. 2016) is considered. The difference between these two values could be due to potential spallation excesses on light Xe isotopes for the latter, but the debate remains open (Avice et al. 2018a; Conrad et al. 2016). In any case, the extent of fractionation is comparable for Mars and Earth. Having a similar isotopic fractionation for atmospheric Xe on Mars and Earth is striking. Indeed, the escape process proposed by Zahnle et al. (2019) depends on parameters such as the gravity of the body, the ionization rate of Xe in the atmosphere and the Solar irradiation activity, and probably the strength and shape of the magnetic field since Xe escapes as ions. All these parameters have different values on Earth and Mars and some of them have also varied with time. It is thus surprising to end up with a similar isotopic fractionation and depletion factor for Mars and Earth atmospheric Xe. If the extent of isotopic fractionation of atmospheric Xe on Earth and Mars is similar, the timing of events might be different. Recently, Cassata (2017) argued that the > 4 Ga ALH84001 and NWA7084 ("Black Beauty") Martian meteorites contain ancient Martian atmospheric gases with little if any Xe isotopic fractionation relative to modern Martian atmospheric Xe. One conclusion of this study is that the isotopic fractionation and escape of Mars atmospheric Xe occurred very early in the planet's history and ceased around 4.4-4.0 Ga (Fig. 6). This observation is in line with studies arguing for an early stage of strong atmospheric escape on Mars starting from the time when a steam atmosphere was outgassed from the magma ocean and lasting until 4 Ga (*e.g.* Lammer et al. (2013)). It is thus tempting to link the escape of xenon with a major transition in the Martian environment and the disappearance of permanent liquid water on Mars by the end of the Noachian (Bibring et al. 2006). As in the case of terrestrial Xe (Zahnle et al 2019), hydrogen from dissociation of atmospheric $H_2O$ could have played a role in the Xe escape and isotopic fractionation, but at different periods for Earth and Mars. It remains to be understood why the extent of Xe isotopic fractionations are comparable for both planetary atmospheres.

**2.1. Martian atmospheric nitrogen**

The present-day isotopic composition of Martian atmospheric $N_2$ measured by the Mars Science Laboratory (Curiosity rover) is $\delta^{15}N = 572\pm90$ ‰ (Wong et al. 2013). The initial Martian value could have been $\delta^{15}N = -30$ ‰ as measured in the Chassigny meteorite (Mathew et al. 1998). Thus, either the Martian atmosphere was contributed by a component rich in $^{15}N$ such as a comet (see Marty et al. 2016), or N isotopes were fractionated during non-thermal nitrogen escape to space (McElroy et al. 1976). The modern nitrogen isotopic composition is consistent with values measured in Martian meteorites, most of them being younger than 1.3 Ga. However, the 4.1 Ga ALH84001 Martian meteorite contains different N components including a presumed Martian atmospheric end-member, identified with Xe isotopes, which is less rich in $^{15}$N. This suggests that the evolution of the N isotope composition by atmospheric escape processes is younger than ~4 Ga. Fig. 6 compares the different evolutions of N and Xe isotope fractionation through time as constrained by the present data. Clearly, the history of the Martian atmosphere is far to be unveiled, and more data for various geological ages are needed.

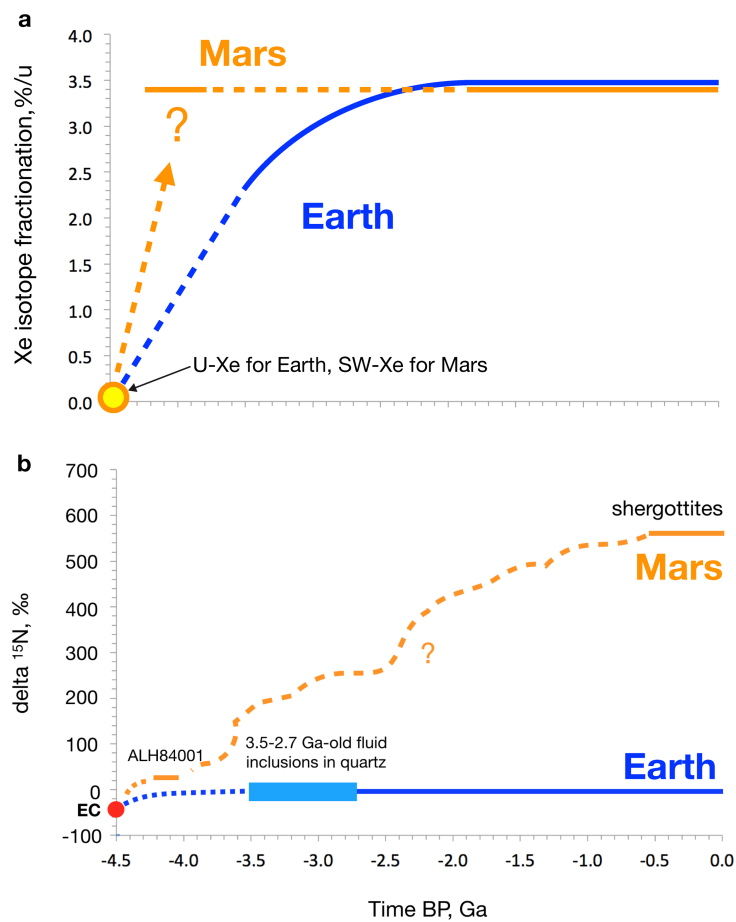

*Figure 6: Comparative evolutions of xenon and nitrogen isotopes as constrained by available data.* a: Xe isotope fractionation through time. The terrestrial ancestor of Xe is thought to be U-Xe, a composite composition resulting from mixing cometary and solar, or chondritic Xe (Marty et al. 2017). For Mars, a solar Xe composition has been advocated (Conrad et al. 2016) and this component is identified in mantle-derived Xe (Mathew et al. 1998). At this scale, considering U-Xe or solar Xe does not make a difference. Xenon in >4 Ga Martian meteorites appears to be` already isotopically fractionated (Cassata 2017), suggesting that Xe isotope fractionation took place early. On Earth, this has been a continuous process still in progress during the Archean (Avice et al. 2018). In the case of nitrogen, the situation appears reverse. The "Old" Martian meteoritic component did not *show $^{15}N$ enrichments* (Mathew and Marti 2001) *whereas "recent" (<1.3 Ga) Martian meteorites show Xe isotope fractionation similar to that of the modern atmosphere* (Mathew et al. 1998). *Martian samples (meteorites, returned samples) with intermediate ages would allow investigating more precisely the long-term evolution of the Martian atmosphere.*

## 3. A desperate lack of data for Venus

The striking similarities between atmospheric Xe on Earth and Mars call for a new point of comparison and the Venusian atmosphere is an ideal candidate here. Comparative planetology was already successful at deciphering the origin and evolution of Earth and Mars. Venus is often depicted as the "sister" planet of Earth. Yet, Venus took a very different evolutionary path that led to an unhabitable world and reasons for this different history are matter of debate. Contrary to the case of Mars, noble gas data for the Venusian atmosphere are lacking and further investigations are mandatory (Chassefière et al. 2012). The current state of knowledge was already reviewed by (Wieler 2002), and data obtained *in-situ* have also been recently reviewed by (Johnson and de Oliveira 2019). Only some major points are highlighted below.

The abundances of Ne and Ar in the Venusian atmosphere are higher than those of the Earth by about one order of magnitude. Even if the Ne/Ar ratio is close to chondritic values, these high abundances of Ne and Ar and a $^{20}Ne/^{22}Ne$ ratio close to 12, higher than the chondritic end-member (about 10.7, Busemann et al. (2000)) suggest a contribution from a solar-derived component to the Venusian atmosphere (Solar Wind Ne implanted on grains (Bogard 1988) or solar neon from the nebula). The Venusian $^{40}Ar/^{36}Ar$ ratio of 1.11±0.02 is lower than the terrestrial ratio of 298.6 (Lee et al. 2006), and this value has been used to argue for incomplete degassing of Venus of its radiogenic $^{40}Ar$ produced by the radioactive decay of $^{40}K$ ($T_{1/2}$ = 1.25 Ga) (Kaula 1999). Based on the atmospheric $^{36}Ar$ abundance, the $^{40}Ar/^{36}Ar$ ratio, the K/U ratios measured by the Venera landers and assuming a terrestrial-like uranium abundance, Kaula (1999) computed that on Venus only 25% of $^{40}Ar$ produced in silicate reservoirs is degassed into the atmosphere. This is half the fraction of $^{40}$Ar degassed on Earth (Allègre et al. 1987). The abundance of $^{84}$Kr in the Venusian atmosphere is 25 (+13/-18) ppb vol. The $^{84}$Kr/$^{36}$Ar elemental ratio suggests a solar-like abundance pattern for heavy noble gases (Fig. 1a). However, there is no data on the isotope composition of Kr to test this hypothesis. The situation is even worse for Xe with only an upper limit of 10 ppb for $^{132}$Xe having been determined. The absence of precise data on the abundance of Xe precludes from evaluating if the abundance of Xe is close to solar values or if there is a "missing Xe" also on Venus. The absence of isotope data also prevents one from checking if the precursor of Venus atmospheric Xe is Chondritic, Solar or U-Xe and if this starting isotopic composition has been mass-dependently fractionated similarly to the case of Earth and Mars.

The current abundance of $N_2$ in the Venus atmosphere is about 3.5 % vol., the remaining of the atmosphere being 96.5 % vol. of $CO_2$ (see the review by Fegley (2003) on the composition of the Venusian atmosphere). Interestingly, Venus has a higher abundance of nitrogen than the Earth in terms of column mass (surface pressure divided by gravitational acceleration) with 23.7 x 10$^3$ kg.m$^{-2}$ of $N_2$ for Venus and 7.8 x 10$^3$ kg.m$^{-2}$ of $N_2$ for the Earth (Catling and Kasting 2017). This difference suggests that, relatively to their respective masses, Venus could have been endowed with more nitrogen than the Earth. Alternatively, some studies suggest that a large fraction of terrestrial nitrogen also resides in silicate reservoirs of the planet (Cartigny and Marty, 2013 and refs. therein). Unfortunately, the isotopic composition of nitrogen in the atmosphere of Venus remains largely unknown with estimates of terrestrial-like value but with an uncertainty of about 20 % covering a wide range of reference values for reservoirs of nitrogen in the solar system (Chassefière et al. 2012; Hoffman et al. 1979). This prevents to compare the cycle of nitrogen on both planets, to estimate which cosmochemical reservoirs contributed this element to Venus and to evaluate if the budget of nitrogen has been altered by atmospheric escape processes.

New space mission efforts toward a measurement of the elemental and isotopic composition of noble gases and nitrogen in the Venus atmosphere are required and proposed to space agencies worldwide (*cf.* Chapter 7). The paucity of noble gases and their inertness prevent remote observations. To our knowledge, three types of investigations could be carried out: i) *in situ* measurement in the Venusian atmosphere at low velocity, e.g., on board a balloon or a vehicle having a slow descent through the atmosphere: ii) sample collection at high velocity, e.g., by a probe skimming into the Venusian atmosphere followed by measurement in space; iii) an atmosphere's sample return. Collecting a sample below the homopause and measuring *in-situ* its noble gas composition in space with a compact probe look promising, especially with the development of a new generation of miniature mass spectrometers (Avice et al. 2019).

## 4. Concluding remarks

Recent studies allowed one to access the composition of the Earth's atmosphere and of the ancient seawater, up to 3.3 Ga. A major result is that the isotopic composition of atmospheric xenon evolved through time, at least until 2 Ga. This evolution was probably linked to a specific Xe escape process taking place in the early anoxic atmosphere under irradiation, and thus ionization, by the young Sun. Nitrogen-argon systematics demonstrates that the partial pressure of nitrogen in the Archean was similar to, or lower than, the modern value. This implies that a higher abundance of nitrogen in the atmosphere than today cannot explain the faint young Sun paradox. The isotopic composition of atmospheric nitrogen was also close to modern values implying little isotopic evolution since the original delivery of nitrogen to Earth. Additionally, analyses of Archean fluid inclusions point toward oceanic waters having salinities in the same range as today if oceans were warmer than present (Robert & Chaussidon, 2006: Tartèse et al., 2016). The atmospheres of Earth and Mars followed distinct paths as evidenced by geochemical studies of Xe and N. The isotope fractionation and potential loss of Xe happened early on Mars, contrary to a protracted evolution for the Earth, but the interplay between the parameters controlling xenon escape and its isotopic evolution remains to be understood. For the Earth, atmospheric nitrogen seems to have suffered little isotopic fractionation while on Mars nitrogen is heavily isotopically fractionated.

The exercise of comparative planetology for the Solar system cannot be completed without considering Venus, the Earth's sister planet. However, too little is known about this planet, especially about the elemental and isotopic composition of noble gases in its atmosphere. Future space missions will certainly unveil some of the remaining mysteries and help to understand the origin and evolution of planetary atmospheres and, ultimately, what makes a planetary body habitable.

**Acknowledgments**

G. A. acknowledges the DIM ACAV+ program (Région Ile-de-France) for its financial support. B.M. acknowledges the European Research Council (grant PHOTONIS 695618 to B.M.).